\documentclass[%
 aip,
 amsmath,amssymb,
 reprint,%
]{revtex4-1}

\usepackage{graphicx}
\usepackage{dcolumn}
\usepackage{bm}

\usepackage[utf8]{inputenc}
\usepackage[T1]{fontenc}
\usepackage{mathptmx}
\usepackage{etoolbox}

\makeatletter
\def\@email#1#2{%
 \endgroup
 \patchcmd{\titleblock@produce}
  {\frontmatter@RRAPformat}
  {\frontmatter@RRAPformat{\produce@RRAP{*#1\href{mailto:#2}{#2}}}\frontmatter@RRAPformat}
  {}{}
}%
\makeatother

\usepackage{multirow}

\begin{document}

\preprint{AIP/123-QED}

\title[]{Self-diffusion and shear viscosity for the TIP4P/Ice water model}
\author{{\L}ukasz. Baran}
\email{lukasz.baran@mail.umcs.pl}
\author{Wojciech R{\.z}ysko}%
\affiliation{Department of Theoretical Chemistry, Institute of Chemical Sciences, Faculty
	of Chemistry, Maria-Curie-Sklodowska University in Lublin, Pl. M
	Curie-Sklodowskiej 3, 20-031 Lublin, Poland}
\author{Luis G. MacDowell}
\affiliation{Departamento de Qu\'{\i}mica-F\'{\i}sica, Facultad de Ciencias Qu\'{\i}micas, Universidad Complutense de Madrid, 28040 Madrid, Spain.}

\date{\today}

\begin{abstract}
With an ever-increasing interest in water properties, 
many intermolecular force fields have been proposed to describe the behavior of
water. Unfortunately, good models for liquid water usually cannot 
provide simultaneously an accurate melting point for ice.
For this reason, the TIP4P/Ice model was developed at targeting 
the melting point, and has become the preferred choice for simulating
ice at coexistence. Unfortunately, available data for
its dynamic properties in the liquid state are scarce.
Therefore, we demonstrate a series of simulations aimed at the calculation of transport coefficients for the TIP4P/Ice model over a large range
of thermodynamic conditions, ranging from $T=245$ K to $T=350$ K for the
temperature and from  $p=0$ to $p=500$ MPa for the pressure.
We have found that the self-diffusion (shear viscosity)
exhibits smaller (increased) values than  TIP4P/2005 and experiments.
However, rescaling the temperature with respect to the triple point temperature
as in a corresponding states plot
we find TIP4P/Ice compares very well with TIP4P/2005  and to experiment.
Such observations allow us to infer that
despite the different original purposes of these two models examined here, one can benefit from
a vast number of reports regarding the behavior of transport coefficients
 for the TIP4P/2005 model and utilize them following the routine described in this paper.  
\end{abstract}

\maketitle

\section{Introduction}

\indent \indent Water is one of the most ubiquitous substances on earth. Hence, it is not surprising that great experimental and computational efforts have been devoted to investigate its properties. Such measurements are of great practical
significance, but they have showed also that water exhibits a large number of interesting anomalies. These range from the well known
negative expansion coefficient below 4$^\circ$ C, to the sharp increase of response functions such as the compressibility upon cooling.  \cite{anomalie} 

This anomalous behavior also extends to dynamic properties and transport coefficients.  Since the end of the 19th century it is known that viscosity of water decreases
with the increase of pressure. \cite{visc_press} On the other hand, the diffusion coefficient increases  as a function of pressure for both translation and rotation.  This unusual behavior of transport properties is related to the breakage of the bond hydrogen network upon compression.  \cite{caupin_visc} Indeed, the conversion between highly ordered tetrahedral water arrangements and disordered domains of high density are in the origin of most of water's anomalies. \cite{cerdeirina22} This equilibrium, which persists in room temperature water,   can be traced back to thermodynamic anomalies of metastable supercooled water. \cite{Debenedetti_2003} Therefore,  the properties of supercooled water have been extensively studied,
including the fragile-to-strong transition, \cite{fragile2strong1, fragile2strong2}
or the preservation of the Stokes-Einstein relation. \cite{caupin, corr_len}

The signature of bulk water's anomalies is also relevant to the study of interfaces. Particularly, the interface of supercooled liquid water at carbon nanotubes \cite{slip_tubes} and slit pores \cite{slip1} exhibits giant slip lengths   which increase upon cooling and appear to be related with the failure of low temperature bulk water to follow the Stokes-Einstein relation. Similarly, the understanding of surface premelting \cite{layering, Llombart_Ice} and interfacial premelting, \cite{qll_solid} as well as the related properties of the resulting quasi-liquid layer (QLL), have been the matter of intense debate. \cite{qll_msd, qll_patrykiejew, qll_louden} For ice premelting, for example, computer simulation studies show enhanced diffusivity of molecules at the water/vapor surface, \cite{qll_louden,qll_msd,weber18} but intriguingly, some experiments claim a strong increase of viscosity of the premelting films relative to bulk values. \cite{murata15,nagata19}

In this regard, \textit{in silico} studies of water properties provide very useful complementary information to experiments. Of course, such effort largely relies on the availability of intermolecular force fields, which are parameterized in such a way to be able to reflect behavior of water as close as possible. Among the most popular point charge models, one can mention TIP4P \cite{tip4p}
and its extended version TIP4P/2005 \cite{Vega2005}; SPC/E, \cite{BerendsenSPCE} TIP5P \cite{JorgensenTIP5P} 
and many other \cite{water_models}.  Unfortunately, it appears that point charge models which are very good at predicting liquid water properties cannot simultaneously provide an accurate melting point for ice \cite{blazquez22}. In order to study the ice and ice-vapor equilibrium properties,
which are of great importance for the determination of growth rates and shapes of snowflakes, \cite{rates}
friction and lubrication of surfaces, \cite{frict} and many other important phenomena, \cite{coex_rev1}         
the TIP4P/Ice model has been developed. \cite{VegaIce}
This force field exhibits a melting temperature of $T\approx270$ K  \cite{CondeMelting}
in ambient conditions which is very close to the experimental result
and also reproduces the melting line. Therefore, it is pivotal to have a comprehensive outlook into the bulk water properties of the TIP4P/Ice model for studies of ice and water coexistence.  However, it is worth emphasizing that while it is easy to find values of transport coefficients for
different water models in a variety of system's conditions, \cite{corr_len, transport_water1, transport_sanz} 
the literature data for TIP4P/Ice model has hardly been explored, except for a limited number of thermodynamic states. \cite{qll_msd, qll_louden, deKonig}

In view of this, the aim of this paper is to calculate both diffusion coefficients and shear viscosities of the TIP4P/Ice model over a large range of thermodynamic conditions, from ambient temperature to the undercooled regime, as well as for a large range of pressures  spanning atmospheric conditions to the  hundreds of MPa.


\section{Methods}\label{sec:method}

To determine the diffusion coefficient we have used the Einstein relation,
which involves the calculation of mean squared displacement (MSD)  of individual water molecules:
\begin{equation}
 \left <\Delta r^2(t) \right > = \left < ({\bf r}(t)-{\bf r}(0))^2 \right >
 \label{msd}
\end{equation}
where ${\bf r}(t)$ is the position of a water molecule at time $t$, and the triangular brackets denote a thermal average over all time origins and individual particles.
For an $n-$dimensional system, the MSD is linear with time
and the slope is related to the diffusion coefficient $D$ as $ \left <r^2(t) \right> =2nDt$. \cite{mcquarrie76}

In order to calculate the viscosity, we have employed the Green-Kubo relation that involves the calculation of autocorrelation functions of components of the stress tensor, as: 
\begin{equation}
 G_{\alpha \beta}=\frac{V}{k_BT}\left < \sigma_{\alpha \beta}(t)\sigma_{\alpha \beta}(0) \right >
\end{equation}
\noindent where $V$ is the volume of the system, $\sigma_{\alpha \beta}$ represents the ${\alpha,\beta}=x, y, z$ component of the stress tensor, and $k_B$ is the Boltzmann constant. \cite{mcquarrie76} To improve the statistics, we do not restrict to
the off-diagonal components of the stress tensor. Daivis and Evans have shown that the diagonal components 
of the stress tensor are larger by the factor of $2-(2/n)$ than the off-diagonal elements, 
where $n$ is the dimensionality of the system. \cite{diag_eta}
Therefore, one can use all of the six components of the stress tensor 
in the calculation of the shear viscosity, just given the fact that they are scaled by the adequate factor.
Knowing that, the shear viscosity is then calculated as 

\begin{equation}
 \eta=\int_0^\infty G_\eta(t) dt
\end{equation}

\noindent where $G_\eta(t)=\frac{1}{6}[G_{xy}+G_{xz}+G_{yz}+\frac{3}{4}(G_{xx}+G_{yy}+G_{zz})]$.

Water was modeled with the use of TIP4P/Ice force field. \cite{VegaIce}
The number of water molecules was equal to 1280 for all systems studied.
We decided to use such system size to avoid system size effects due to insufficient
number of molecules and yet being able to perform simulations in a reasonable time
 which is particularly important in the case of the shear viscosity
as these calculations are computationally expensive.
Molecular dynamics simulations of bulk water were performed using the LAMMPS package. \cite{LAMMPS}
Trajectories were evolved using the velocity-Verlet algorithm with a 2 fs time step. 
Bonds and bond angles were constrained by the use of the SHAKE algorithm. 
Both the temperature and pressure was set using Nose-Hoover chains algorithms \cite{nhchains, nhchains2} 
with damping factor equal $\tau=2$ ps and the number of chains $M=3$. 
To remain consistent with the definition of TIP4P/Ice, 
all dispersion interactions were truncated at $8.5$ \AA.
Long-range coulombic interactions were computed using the particle-particle
particle-mesh method. \cite{p3m} 
The charge structure factors were evaluated with the fourth-order interpolation scheme and 
a grid spacing of $1$~\AA, resulting in 36, 32 and 36 vectors in the x, y, z directions
in reciprocal space, respectively.

\begin{figure}[t!]
\begin{center}
 \includegraphics[width=0.5\textwidth]{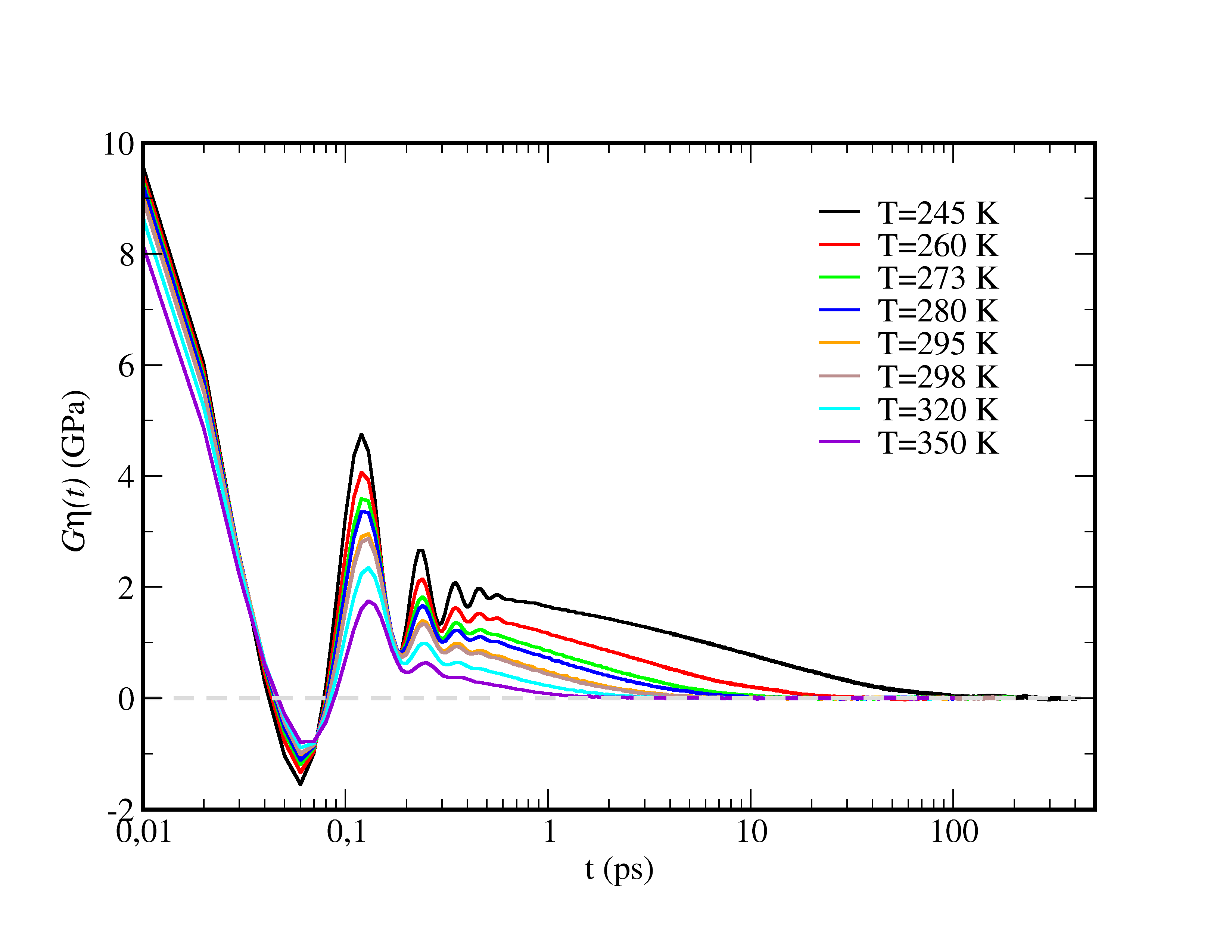}
 \caption{Stress autocorrelation functions $G_{\eta}(t)$ evaluated at different temperatures in ambient pressure, $p=0$ MPa. }
 \label{GK_temps} 
\end{center}
\end{figure}

The simulation scheme involved two steps. In the first one, 
simulations for bulk water have been performed for $15$ ns in the $NpT$ ensemble, in order to obtain an accurate average density.
Then, the system was accordingly rescaled to the given average density, and further $40$ ns runs 
in $NVT$ ensemble were performed in which the relevant trajectories for the calculation of transport coefficients
have been gathered over the last 30 ns,  every 15 ps. For the calculation of shear viscosities, the
stress tensor components have been printed every 10 fs. 
Simultaneous evaluation of the diffusion coefficients and shear viscosity
has a particular advantage that one can test their coupling right away
using a single molecular dynamics simulation.

Following the above simulation scheme, 
the translational mean square displacement and shear viscosities were calculated at eight different temperatures 
spanning from $T=245$ K to $T=350$ K in the pressure range of $p=0-500$ MPa.
In all the conditions studied, the linear regime of MSD has been established
starting from the first saved configuration. 
However, in the case of the Green-Kubo autocorrelation functions, the numerical integration
by trapezoidal rule had to be performed up to the upper limit, denoted as $\tau_{\alpha}$.
Its value has been estimated by the inspection of a characteristic timescale where the autocorrelation function
smoothly decays to 0, so that the contributions from the long tail to the integral
which are a subject to the random noise are omitted. 
Averaged Green-Kubo autocorrelation functions evaluated at different temperatures in ambient pressure 
can be found in Figure~\ref{GK_temps}. 
One can see that there is a significant change in the characteristic timescale $\tau_{\alpha}$,
which increases upon cooling and reaches up to 200 ps for the lowest temperature (cf. Table~\ref{tab1}
in Appendix~\ref{appA}).


\section{Results}

\begin{figure}[htb!]
\begin{center}
 \includegraphics[width=0.5\textwidth]{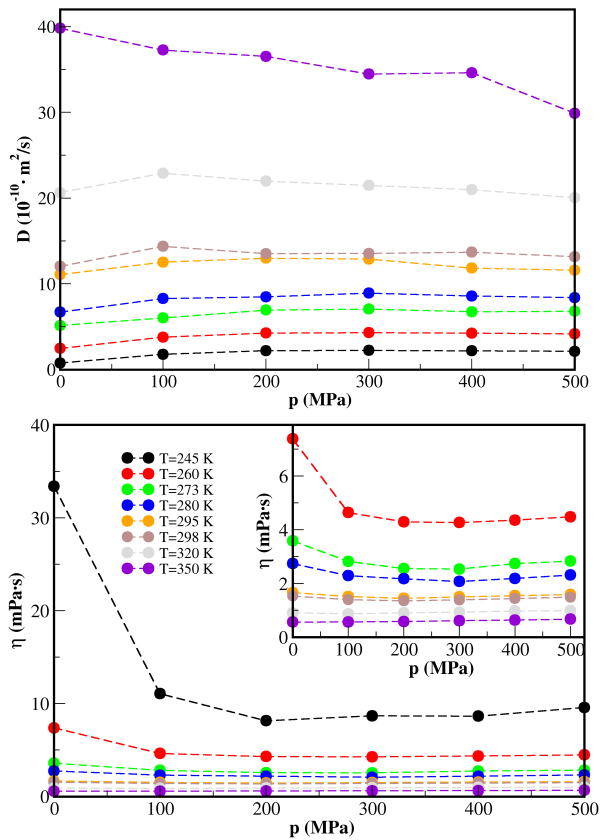}
 \caption{Relation of self-diffusion coefficient (top) and viscosity (bottom) with respect to
 pressure. Labels are the same for both panels. Inset to the bottom part displays
 the magnified data.}
 \label{ice_raw} 
\end{center}
\end{figure}

Isotherms for a given transport coefficient examined in 
the paper are presented in Figure~\ref{ice_raw}
and also can be found in Table~\ref{tab1} shown in Appendix~\ref{appA}.
Let us now compare currently obtained results with the available experimental data. 
For instance, at ambient pressure with this model we obtain $D=12.368~\cdot10^{-10}\mathrm{(m^2s^{-1})}$ and $D=5.079~\cdot10^{-10}\mathrm{(m^2s^{-1})}$, compared to experimental values of $D_{exp}=23~\cdot10^{-10}\mathrm{(m^2s^{-1})}$ \cite{diff_expT298} and $D_{exp}=10.9~\cdot10^{-10}\mathrm{(m^2s^{-1})}$ \cite{diff_expT273} at $T=298$ K and $T=273$ K, respectively. 
One can see that the values of self-diffusion coefficients are lower than those reported experimentally.
On the other hand, the viscosities of liquid water are higher than experimental values \cite{viscosity_exp}
which is not surprising due to the Stokes-Einstein relation. 
Consequently, the aforementioned observations concern all other points studied. 
It is worth highlighting that the decreased (increased) value of
diffusion coefficient (shear viscosity) has already been noted and is consistent with other papers. \cite{water_models, qll_msd}

\begin{figure}[tb!]
\begin{center}
 \includegraphics[width=0.5\textwidth]{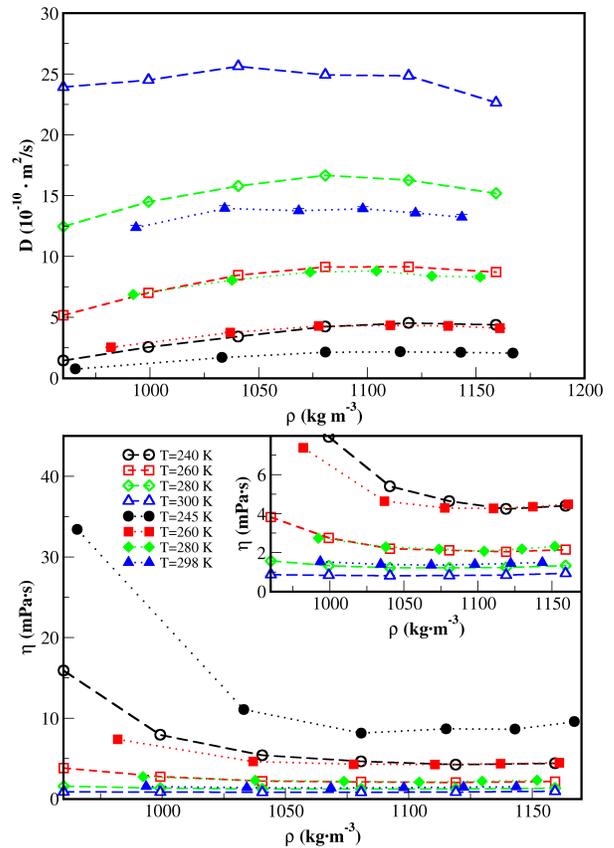}
 \caption{Relation of self-diffusion coefficient (top) and viscosity (bottom) with respect to
 density. Filled and open symbols depict currently obtained data for TIP4P/Ice and
 TIP4P/2005 from Ref.~\cite{transport_sanz}, respectively. Labels are the same for both panels. 
 Inset to the bottom part displays the magnified data.}
 \label{ice_2005_comparison} 
\end{center}
\end{figure}

In view of this fact, it is well-known that TIP4P/2005 is widely used due to its 
successful description of liquid water properties, consistent with experimental data.
Therefore, it seems reasonable to validate our current simulation results with those
obtained for TIP4P/2005 model. 
Figure~\ref{ice_2005_comparison} shows a comparison between our simulations
and those performed by Montero de Hijes \textit{et al}. \cite{transport_sanz} 
Indeed, one can see the same behavior for the transport coefficients  
that, as stated before,  TIP4P/Ice exhibits
a decreased (increased) value of diffusion coefficient (shear viscosity) compared
to TIP4P/2005 water model in the whole range of points studied.
We conjecture that this behavior can be congruent with the shift in
the melting point ($T_m$) of ice Ih and triple points ($T_t$) of these models. 
They were reported to be at $T_{2005,m}\approx 250$ K \cite{CondeMelting} and $T_{2005,t}=252.1$ K \cite{triple_point}
and $T_{ice,m}\approx 270$ K \cite{CondeMelting} and $T_{ice,t}=272.2$ K \cite{triple_point}
for TIP4P/2005 and TIP4P/Ice, respectively. 

\begin{figure}[tb!]
\begin{center}
 \includegraphics[width=0.55\textwidth]{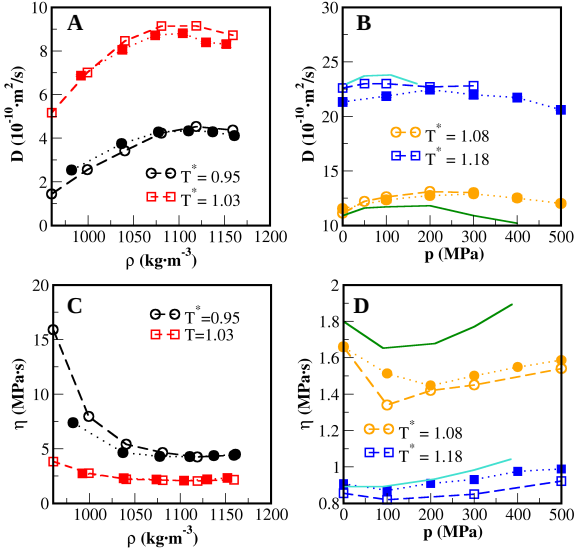}
 \caption{Relation of self-diffusion coefficient (A, B) and viscosity (C, D) with respect to
 density or pressure. Filled and open symbols depict currently obtained data for TIP4P/Ice and
 TIP4P/2005 from Ref.~\cite{transport_sanz} (parts A, C) and Ref.~\cite{transport_water1} (parts B, D), respectively. 
 Solid lines correspond to the experimental data taken from 
 Ref.~\cite{diff_expT298}, Ref.~\cite{diff_expT273}, and Ref.~\cite{viscosity_exp}.}
 \label{reduced_temps} 
\end{center}
\end{figure}

Therefore, a naturally arising questions is whether the origin of these differences
is just due to the temperature shifts? Or as the models were parametrized to describe
different features of water, exhibiting different transport properties is a natural consequence?
Here, we aim to answer these questions. In order to do that, we assumed that 
the simulation points for different models should be presented in  reduced units.
We mapped the points by the law of corresponding states reduced by the triple points of TIP4P/2005 and TIP4P/Ice
models, i.e. $T^*=T_{X}/T_{X, t}$ where $X$ is either TIP4P/2005 or TIP4P/Ice. 
It has to be emphasized that the ratio $T_m/T_c$ between melting to critical points of these two models is
equal to 0.39 and 0.383 for TIP4P2005 and TIP4P/Ice, respectively (see Table V 
of Ref.~\cite{triple_point} for critical points estimation). 
In other words, there is no significant difference whether we choose
to rescale the temperature by a triple or critical point (aside from the temperature scale)
to preserve the universal features shown later on.
The results can be found in Figure~\ref{reduced_temps} and Table~\ref{tab2},~\ref{tab3} in Appendix~\ref{appA}. 

\begin{figure}[tb!]
\begin{center}
 \includegraphics[width=0.45\textwidth]{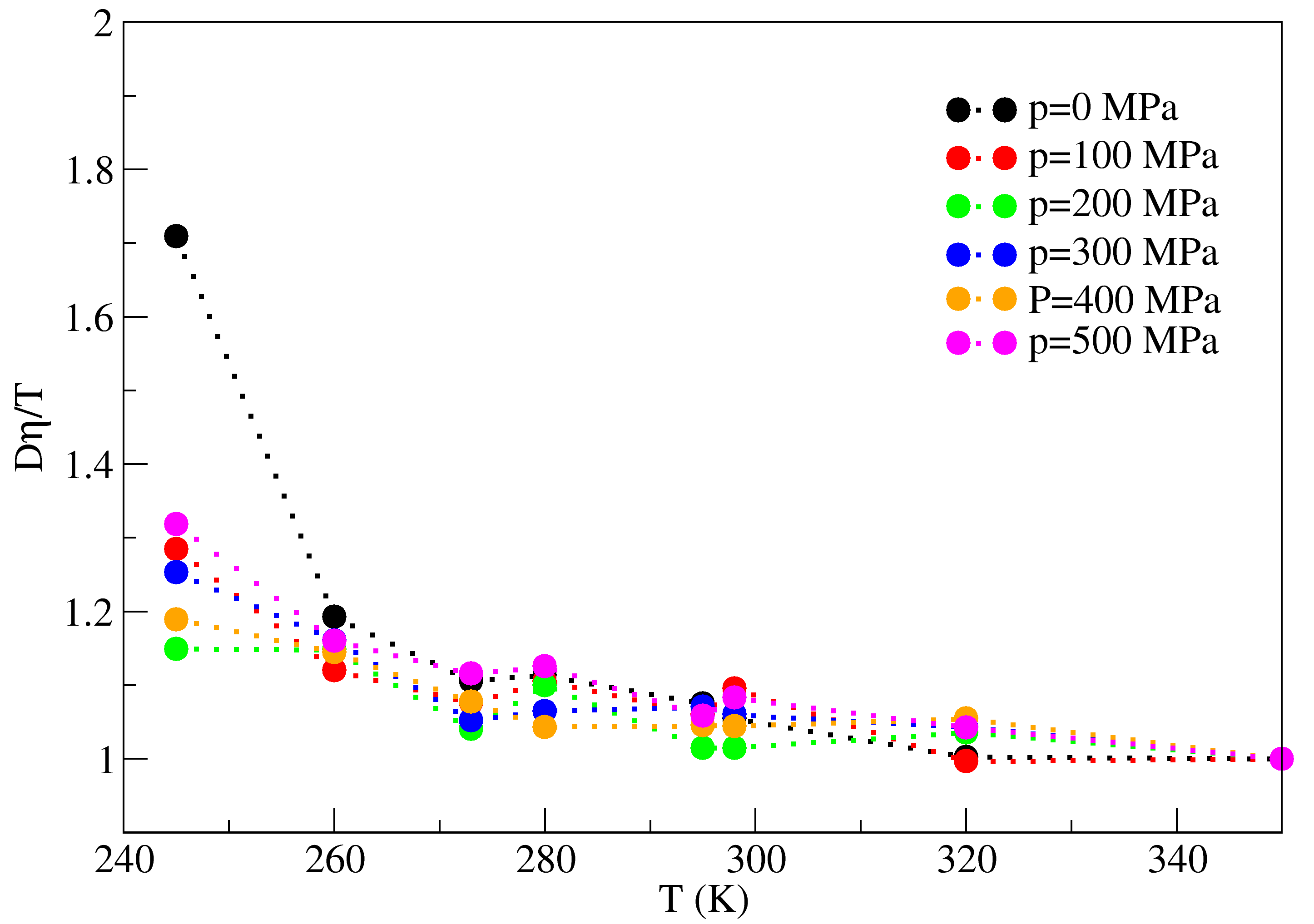}
 \caption{Relation of $D\eta/T$ with temperature, normalized at $T=350$ K for all pressures studied.}
 \label{SE_rel} 
\end{center}
\end{figure}

One can now see an excellent agreement between the two models when expressed in reduced units.
Such observation refers both to the self-diffusion and shear viscosities.
Although one can easily notice that there is a mismatch in the comparison
between current simulation results and the points extracted from Montero de Hijes \textit{et al.} \cite{transport_sanz}
(Fig.~\ref{reduced_temps} A, C) the trends are still preserved within the whole range of points studied.
Differences in the examined densities result from different approaches between our and their
paper. In other words, we wanted to examine the behavior in a wide pressure range for TIP4P/Ice model
rather than explicitly match points from other papers.
Nevertheless, we can observe nearly perfect agreement between two models and also experimental data.
It has to be noted that while for viscosity it seems that the differences are quite large,
they in fact differ by just about 10\%, however still preserving the trend observed experimentally.
It now seems to be clear that the answer to the origin of the differences in transport coefficients
between TIP4P/2005 and TIP4P/Ice models can be attributed to the different
absolute temperature scales. In practice, for comparison with experiments this
means that transport properties obtained at temperature $T$ from 
the TIP4P/Ice model are good estimates for liquid water properties at 
an effective temperature of $T_{ef}=\frac{T_{t,2005}}{T_{t,Ice}} T$.

\begin{figure}[b!]
\begin{center}
 \includegraphics[width=0.5\textwidth]{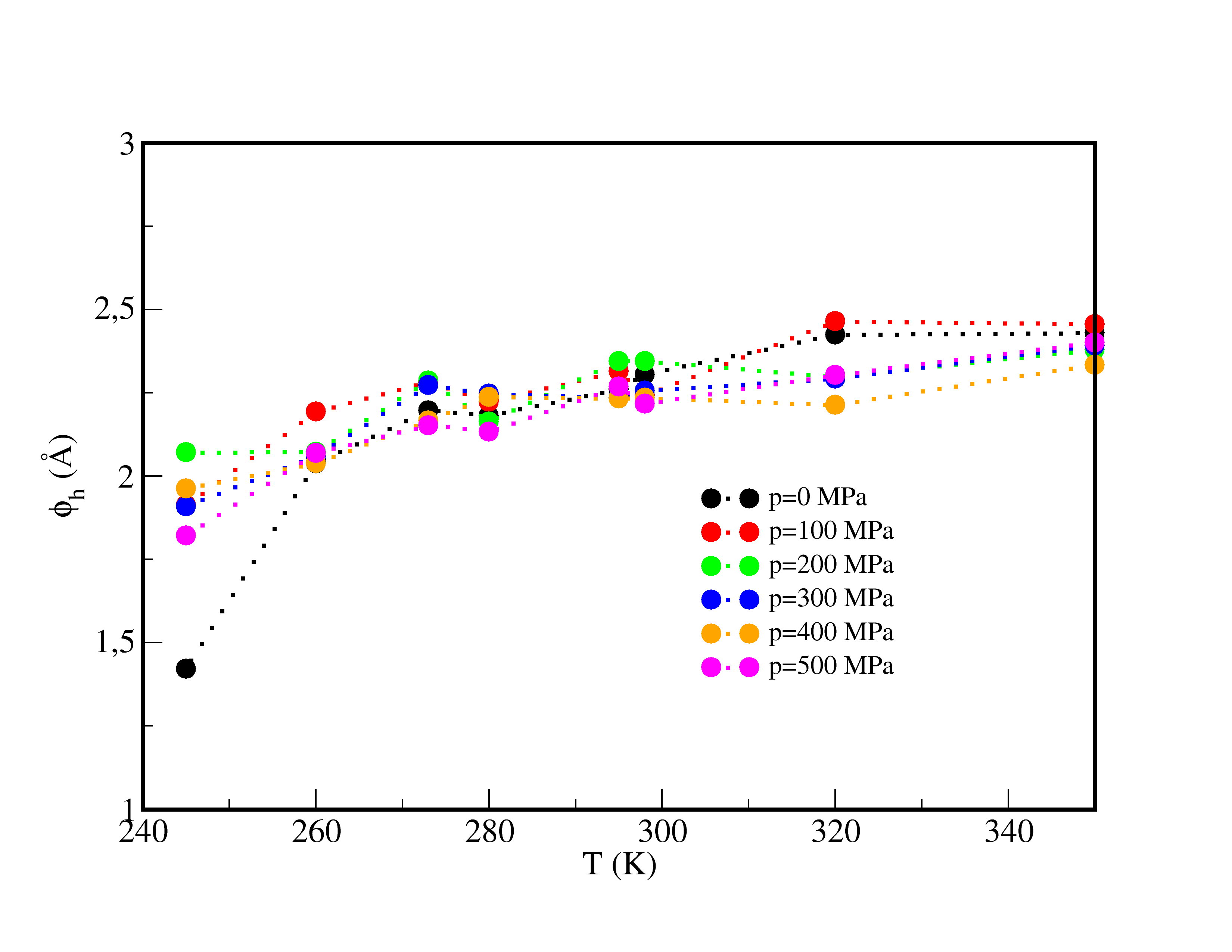}
 \caption{Temperature dependence of hydrodynamic diameter $\phi_h$ for all pressures studied.}
 \label{hydro_diam} 
\end{center}
\end{figure}

According to the Stokes-Einstein relation, the self diffusion of a spherical particle can be related to the viscosity of the surrounding media as $D=k_BT / 3\pi\eta a$, where
$a$ is the molecular diameter. Surprisingly, there exists ample evidence that this hydrodynamic result also holds quite accurately down to molecular scales. In analogy to other water models and experimental systems we checked the Stokes-Einstein (SE) relation for the TIP4P/Ice model by plotting $D\eta/T$ as a function of temperature. The results are
presented in Figure~\ref{SE_rel}.
To show the temperature variation of $D\eta/T$, it has been normalized
by the reference temperature which usually is the
highest temperature examined. In our case, it is taken to be $T=350$ K at all pressures studied. 
We can see that at pressures higher than $p=200$ MPa, the deviations start to  occur already at $T=320$ K, and at a somewhat lower temperature of 
$T=298$ K for all the remaining pressures. This is consistent
with previous work reporting the violation of the SE relation for temperatures
way higher than in regular supercooled liquids which usually is $1.3T_g$ where $T_g$ is the vitrification temperature
(for water $T_g\approx136$ K).  
On the other hand, it has to be noted that aside from the lowest temperature studied $T=245$ K, the deviation is
still moderate and does not exceed 30\%. Importantly, the deviations are roughly independent on the pressure.  Therefore, the SE relation does not serve as a quantitative theory, but it is still useful as an order of magnitude estimate of the viscosity for the TIP4P/Ice model dow to $T=245$~K.
Given that, one can obtain values of viscosity
from the self-diffusion coefficients using Stokes-Einstein 
relation $\eta=\frac{k_B T}{3\pi D \mathrm{a}}$ within an acceptable error margin
with $a=3.1$ \AA~being the molecular diameter of water. 

Since the SE relation is violated, having calculated the values of self-diffusion coefficient and shear viscosity
for a series of isotherms, it is also interesting to check what is the hydrodynamic diameter $\phi_h$ required
to obey the SE relation exactly. It can be extracted as follows:

\begin{equation}
 \phi_h=\frac{k_B T}{3\pi\eta D}
 \label{eq:SE}
\end{equation}

\noindent Figure~\ref{hydro_diam} shows $\phi_h$ calculated from the simulation data. 
Compared to the Lennard-Jones parameter for interaction between the oxygen sites of two
molecules in TIP4P/Ice which is $\sigma_O=3.1668$ \AA~ it exhibits a smaller
value at all temperatures studied. $\phi_h$ is independent on the pressure
and decreases with temperature, however the deviations from $\sigma_O$ does not exceed 30\%
except for the outlier at $T=245$ K at ambient pressure. 
%
%
This feature of decreasing $\phi_h$ upon cooling has been observed in many fragile glassformers. \cite{glassformers}

\section{Conclusions}

In this paper we computed transport coefficient of bulk liquid water using TIP4P/Ice model
at eight different temperatures ranging from $T=245$ K to $T=350$ K in a pressure range of $p=0-500$ MPa.
We emphasize that to our knowledge, such a systematic study for this particular water model has not been performed to date.
We have shown that the self-diffusion (shear viscosity) exhibits decreased (increased) values than experimental findings or
those extracted from TIP4P/2005 water model.
However, if one would switch to the reduced units formalism with respect to the model's triple point (or critical point),
the behavior changes dramatically. In such a case, models are now in excellent agreement with each other and also
reflect experimental data nearly quantitatively.
Such observations allow us to infer that in spite of different original purpose of considered models here,
one can benefit from a vast number of reports regarding the behavior of transport coefficients 
for TIP4P/2005 model and utilize them following the routine described in this paper. 
It seems that the answer to the origin of the differences in transport coefficients
between models is just because of different absolute temperature scales. 

Despite that, we would like to emphasize that we did not study deeply supercooled regime
and therefore, cannot guarantee that this is a universal feature so a special caution has to be paid.
Moreover, in view of these findings, a naturally arising questions are: does the same behavior
would regard other quantities such as  dielectric constant \cite{dielectric_comp},
surface tension \cite{stension_comp}, densities of ices, and perhaps other \cite{comparison_properties}?
Is this the universal feature for all water models? Our results suggest the analogy could be exploited at least for the study of transport properties, so that accurate data for a model with unknown transport coefficients could be inferred from other models with known transport properties. 

\begin{acknowledgments}
   LGM would like to thank E. Sanz for helpful discussions. LGM  would also
   like to acknowledge funding from the Spanish Agencia Estatal de 
   Investigaci\'on under grant PIP2020-115722GB-C21
\end{acknowledgments}

\section*{Data Availability Statement}

The data that support the findings of this study are available from the corresponding author upon reasonable request.

\appendix 
\section{Simulation data}\label{appA}

All of the simulation points are presented in Table~\ref{tab1} with the uncertainities presented in the form
of standard errors. The calculation of uncertainities for the viscosity was straightforward
as we obtained six independent integrals of autocorrelation functions of the traceless 
stress tensor elements. In the case of self-diffusion coefficients, we have followed the same routine
as in the reference. \cite{transport_sanz} Briefly, the trajectory was divided
into four blocks of the same duration equal to 7.5 ns and for each of them
the self-diffusion coefficient was calculated as the slope of the mean-squared displacement
as explained in Section~\ref{sec:method}.

Tables~\ref{tab2} and \ref{tab3} show the comparison of the transport properties of liquid water
for TIP4P/2005 and TIP4P/Ice water models in the reduced temperatures. 
As described in the text, it shows an excellent agreement between
the models as well as with experimental findings.

\begin{table}[htb!]
\begin{center}
\begin{tabular} {|c|c|c|c|c|c|}
 \hline
 \hline
 T (K) & p (MPa) & $\rho$ (kg$\cdot$m$^{-3}$) & D (10$^{-10}\cdot$m$^2$s$^{-1}$) & $\eta$ (mPa$\cdot$s) & $\tau_{\alpha}$ (ps)\\
\hline
 \multirow{6}{*}{245} 
 & 0 & 965.62    & 0.756 (0.013) & 33,392 (2.890)  & 200 \\
 & 100 & 1033.12 & 1.696 (0.020)  & 11.069 (0.186) & 50 \\
 & 200 & 1080.76 & 2.126 (0.016) & 8.151 (0.17)    & 30 \\
 & 300 & 1115.19 & 2.166 (0.019) & 8.681 (0.224)   & 35 \\
 & 400 & 1143.04 & 2.117 (0.022) & 8.641 (0.149)   & 37.5\\
 & 500 & 1167.10 & 2.058 (0.063) & 9.578 (0.192)   & 50 \\ 
\hline      
 \multirow{6}{*}{260} 
 & 0   & 982.01  & 2.533 (0.032) & 7.379 (0.197)  & 55\\
 & 100 & 1036.89 & 3.744 (0.041) & 4.638 (0.084)  & 30 \\
 & 200 & 1077.63 & 4.283 (0.036) & 4.291 (0.137)  & 27.5 \\
 & 300 & 1110.59 & 4.334 (0.064) & 4.265 (0.083)  & 25 \\
 & 400 & 1137.21 & 4.291 (0.089) & 4.355 (0.098)  & 20 \\
 & 500 & 1161.02 & 4.110 (0.050) & 4.479 (0.070)  & 17.5 \\  
 \hline
  \multirow{6}{*}{273}
 & 0   & 989.63  & 5.079 (0.096) & 3.585 (0.076)  & 20\\
 & 100 & 1037.72 & 6.212 (0.030) & 2.821 (0.0604) & 20\\
 & 200 & 1075.34 & 6.845 (0.039) & 2.555 (0.044)  & 10\\
 & 300 & 1106.45 & 6.946 (0.063) & 2.533 (0.019)  & 10\\
 & 400 & 1132.31 & 6.742 (0.060) & 2.739 (0.033)  & 10 \\
 & 500 & 1155.31 & 6.561 (0.036) & 2.832 (0.035)  & 10\\  
 \hline
 \multirow{6}{*}{280}
 & 0   & 992.24  & 6.866 (0.040) & 2.738 (0.069) & 15\\
 & 100 & 1037.71 & 8.054 (0.125) & 2.292 (0.011) & 15\\
 & 200 & 1073.73 & 8.717 (0.095) & 2.175 (0.031) & 15\\
 & 300 & 1104.23 & 8.810 (0.102) & 2.072 (0.029) & 10\\
 & 400 & 1129.72 & 8.392 (0.049) & 2.185 (0.017) & 10 \\
 & 500 & 1151.99 & 8.312 (0.148) & 2.313 (0.013) & 10 \\  
  \hline
 \multirow{6}{*}{295}
 & 0   & 993.95  & 11.528 (0.231) & 1.658 (0.008) & 10 \\
 & 100 & 1035.53 & 12.341 (0.283) & 1.513 (0.011) & 8\\
 & 200 & 1069.76 & 12.734 (0.169) & 1.447 (0.016) & 7 \\
 & 300 & 1098.51 & 12.892 (0.027) & 1.500 (0.014) & 7 \\
 & 400 & 1123.58 & 12.501 (0.283) & 1.549 (0.022) & 6 \\
 & 500 & 1145.33 & 12.009 (0.043) & 1.586 (0.014) & 7 \\  
 \hline
 \multirow{6}{*}{298}
 & 0   & 993.51  & 12.368 (0.198) & 1.532 (0.024) & 8 \\
 & 100 & 1034.35 & 13.949 (0.228) & 1.396 (0.011) & 8\\
 & 200 & 1068.45 & 13.752 (0.207) & 1.354 (0.017) & 7\\
 & 300 & 1097.94 & 13.922 (0.162) & 1.390 (0.019) & 7\\
 & 400 & 1122.17 & 13.564 (0.126) & 1.441 (0.013) & 6\\
 & 500 & 1143.57 & 13.236 (0.219) & 1.488 (0.014) & 7 \\  
  \hline
 \multirow{6}{*}{320}
 & 0   & 990.33  & 21.317 (0.194) & 0.907 (0.011) & 5\\
 & 100 & 1028.62 & 21.856 (0.132) & 0.870 (0.009) & 4\\
 & 200 & 1061.28 & 22.447 (0.191) & 0.909 (0.004) & 4\\
 & 300 & 1089.26 & 21.983 (0.350) & 0.930 (0.007) & 4\\
 & 400 & 1113.14 & 21.723 (0.059) & 0.975 (0.012) & 4\\
 & 500 & 1134.37 & 20.598 (0.069) & 0.988 (0.008) & 3 \\ 
 \hline
  \multirow{6}{*}{350}
 & 0   & 979.43  & 37.512 (1.116) & 0.562 (0.006) & 4 \\
 & 100 & 1017.62 & 36.706 (0.250) & 0.569 (0.005) & 2.5\\
 & 200 & 1049.17 & 36.923 (0.426) & 0.584 (0.001) & 2.5\\
 & 300 & 1075.84 & 34.953 (0.279) & 0.613 (0.002) & 2.5\\
 & 400 & 1099.59 & 34.474 (0.232) & 0.637 (0.004) & 2\\
 & 500 & 1120.75 & 31.801 (0.303) & 0.672 (0.003) & 2 \\ 
 \hline
 \hline
\end{tabular}
\caption{Simulation results of TIP4P/Ice water model for the shear viscosity and diffusion coefficients.
$\tau_\alpha$ depicts the upper limit in the integral over the autocorrelation functions. 
The standard errors are given in the parentheses.}
\label{tab1}
\end{center}
\end{table}

\begin{table}
\begin{center}
\scalebox{0.85}{
 \begin{tabular}{|c|c|c|c|c|c|c|c|c|}
 \hline
 \hline
 $T^*$&  p$_{\mathrm{TIP4P/2005}}$ & p$_\mathrm{TIP4P/Ice}$ & p$_\mathrm{expt.}$ & $T_{\mathrm{TIP4P/2005}}$ & $T_{\mathrm{TIP4P/Ice}}$ & $D_\mathrm{{TIP4P/2005}}$ & $D_\mathrm{{TIP4P/Ice}}$
 & $D_{\mathrm{expt.}}$  \\
 \hline 
 \multirow{4}{*}{1.08} & 0.1  & 0    & 0.1 & 273 & 295 & 11.1  & 11.53 & 10.9  \\
                        & 100  & 100  & 100 & 273 & 295 & 12.61 & 12.34 & 11.7 \\
                        & 200  & 200  & 200 & 273 & 295 & 13.1  & 12.73 & 11.8 \\
                        & 300  & 300  & 300 & 273 & 295 & 13.   & 12.89 & 10.9 \\
  \hline  
   \multirow{4}{*}{1.18} & 0.1 & 0   & 10 & 298 & 320 & 22.60 & 21.32 & 23.0 \\
                          & 100 & 100 & 110 & 298 & 320 & 23.   & 21.86 & 23.8 \\
                          & 200 & 200 & 170 & 298 & 320 & 22.7  & 22.45 & 23.  \\
                          & 300 & 300 &     & 298 & 320 & 22.8  & 21.98 &      \\
  \hline
  \hline
 \end{tabular}}
\end{center}
\caption{Comparison of simulation results for two water models for the self-diffusion coefficient $D$ in $10^{-10}$ m$^2\cdot$s$^{-1}$.
Experimental temperature is the same as for TIP4P/2005. Experimental values are taken from Ref.~\cite{diff_expT298} and Ref.~\cite{diff_expT273}
whereas TIP4P/2005 are extracted from Ref.~\cite{diff_2005}. }
\label{tab2}
\end{table}

\begin{table}
\begin{center}
\scalebox{0.85}{
 \begin{tabular}{|c|c|c|c|c|c|c|c|c|}
 \hline
 \hline
 $T^*$&  p$_{\mathrm{TIP4P/2005}}$ & p$_\mathrm{TIP4P/Ice}$ & p$_\mathrm{expt.}$ & $T_{\mathrm{TIP4P/2005}}$ & $T_{\mathrm{TIP4P/Ice}}$ & $\eta_\mathrm{{TIP4P/2005}}$ & $\eta_\mathrm{{TIP4P/Ice}}$
 & $\eta_{\mathrm{expt.}}$  \\
 \hline 
 \multirow{5}{*}{1.08} & 0.  & 0    & 0.1   & 273 & 295 & 1.66  & 1.66 & 1.799  \\
                        & 100  & 100 & 90.7  & 273 & 295 & 1.34 & 1.51 & 1.653 \\
                        & 200  & 200 & 209.6 & 273 & 295 & 1.42 & 1.45 & 1.678 \\
                        & 300  & 300 & 300.4 & 273 & 295 & 1.45 & 1.50 & 1.771 \\
                        & 500  & 500 &       & 273 & 295 & 1.54 & 1.59 & \\
  \hline  
   \multirow{4}{*}{1.18} & 0. & 0    &  0.1   & 298 & 320  & 0.855 & 0.907 & 0.892 \\
                          & 100 & 100 &  90.7  & 298 & 320 & 0.819 & 0.870 &  0.891 \\
                          & 200 & 200 &  209.6 & 298 & 320 & 0.830 & 0.909 &  0.933 \\
                          & 300 & 300 &  300.4 & 298 & 320 & 0.85  & 0.930 &  0.983  \\
                          & 500 & 500 &        & 298 & 320 & 0.922 & 0.988 &      \\
  \hline
  \hline
 \end{tabular}}
\end{center}
\caption{Comparison of simulation results for two water models for the shear-viscosity $\eta$ in mPa$\cdot$s. 
Experimental temperature is the same as for TIP4P/2005.
Experimental values are taken from Ref.~\cite{viscosity_exp} whereas simulation points for TIP4P/2005 are taken from Ref.~\cite{diff_2005}
and Ref.~\cite{transport_water1}.}
\label{tab3}
\end{table}

\clearpage 
\bibliographystyle{ieeetr}
\bibliography{manuscript}

\begin{thebibliography}{10}

\bibitem{anomalie}
P.~Gallo, K.~Amann-Winkel, C.~A. Angell, M.~A. Anisimov, F.~Caupin,
  C.~Chakravarty, E.~Lascaris, T.~Loerting, A.~Z. Panagiotopoulos, J.~Russo,
  J.~A. Sellberg, H.~E. Stanley, H.~Tanaka, C.~Vega, L.~Xu, and L.~G.~M.
  Pettersson, ``Water: A tale of two liquids,'' {\em Chemical Reviews},
  vol.~116, no.~13, pp.~7463--7500, 2016.
\newblock PMID: 27380438.

\bibitem{visc_press}
W.~C. Röntgen, ``Ueber den einfluss des druckes auf die viscosität der
  flüssigkeiten, speciell des wassers,'' {\em Annalen der Physik}, vol.~258,
  no.~8, pp.~510--518, 1884.

\bibitem{caupin_visc}
L.~P. Singh, B.~Issenmann, and F.~Caupin, ``Pressure dependence of viscosity in
  supercooled water and a unified approach for thermodynamic and dynamic
  anomalies of water,'' {\em Proceedings of the National Academy of Sciences},
  vol.~114, no.~17, pp.~4312--4317, 2017.

\bibitem{cerdeirina22}
C.~A. Cerdeiriña, ``Water’s unusual thermodynamics in the realm of physical
  chemistry,'' {\em J. Phys. Chem. B}, vol.~126, no.~35, pp.~6608--6613, 2022.
\newblock PMID: 36001372.

\bibitem{Debenedetti_2003}
P.~G. Debenedetti, ``Supercooled and glassy water,'' {\em Journal of Physics:
  Condensed Matter}, vol.~15, pp.~R1669--R1726, oct 2003.

\bibitem{fragile2strong1}
K.~Ito, C.~T. Moynihan, and C.~A. Angell, ``Thermodynamic determination of
  fragility in liquids and a fragile-to-strong liquid transition in water,''
  {\em Nature}, vol.~398, pp.~492--495, Apr 1999.

\bibitem{fragile2strong2}
R.~Shi, J.~Russo, and H.~Tanaka, ``Origin of the emergent fragile-to-strong
  transition in supercooled water,'' {\em Proceedings of the National Academy
  of Sciences}, vol.~115, no.~38, pp.~9444--9449, 2018.

\bibitem{caupin}
A.~Dehaoui, B.~Issenmann, and F.~Caupin, ``Viscosity of deeply supercooled
  water and its coupling to molecular diffusion,'' {\em Proceedings of the
  National Academy of Sciences}, vol.~112, no.~39, pp.~12020--12025, 2015.

\bibitem{corr_len}
T.~Kawasaki and K.~Kim, ``Identifying time scales for violation/preservation of
  stokes-einstein relation in supercooled water,'' {\em Science Advances},
  vol.~3, no.~8, p.~e1700399, 2017.

\bibitem{slip_tubes}
M.~Majumder, N.~Chopra, R.~Andrews, and B.~J. Hinds, ``Enhanced flow in carbon
  nanotubes,'' {\em Nature}, vol.~438, pp.~44--44, Nov 2005.

\bibitem{slip1}
C.~Herrero, G.~Tocci, S.~Merabia, and L.~Joly, ``Fast increase of nanofluidic
  slip in supercooled water: the key role of dynamics,'' {\em Nanoscale},
  vol.~12, pp.~20396--20403, 2020.

\bibitem{layering}
P.~Llombart, E.~G. Noya, D.~N. Sibley, A.~J. Archer, and L.~G. MacDowell,
  ``Rounded layering transitions on the surface of ice,'' {\em Phys. Rev.
  Lett.}, vol.~124, p.~065702, Feb 2020.

\bibitem{Llombart_Ice}
P.~Llombart, E.~G. Noya, and L.~G. MacDowell, ``Surface phase transitions and
  crystal habits of ice in the atmosphere,'' {\em Science Advances}, vol.~6,
  no.~21, p.~eaay9322, 2020.

\bibitem{qll_solid}
V.-M. Nikiforidis, S.~Datta, M.~K. Borg, and R.~Pillai, ``Impact of surface
  nanostructure and wettability on interfacial ice physics,'' {\em The Journal
  of Chemical Physics}, vol.~155, no.~23, p.~234307, 2021.

\bibitem{qll_msd}
T.~Kling, F.~Kling, and D.~Donadio, ``Structure and dynamics of the
  quasi-liquid layer at the surface of ice from molecular simulations,'' {\em
  The Journal of Physical Chemistry C}, vol.~122, no.~43, pp.~24780--24787,
  2018.

\bibitem{qll_patrykiejew}
M.~M. Conde, C.~Vega, and A.~Patrykiejew, ``The thickness of a liquid layer on
  the free surface of ice as obtained from computer simulation,'' {\em The
  Journal of Chemical Physics}, vol.~129, no.~1, p.~014702, 2008.

\bibitem{qll_louden}
P.~B. Louden and J.~D. Gezelter, ``Why is ice slippery? simulations of shear
  viscosity of the quasi-liquid layer on ice,'' {\em The Journal of Physical
  Chemistry Letters}, vol.~9, no.~13, pp.~3686--3691, 2018.
\newblock PMID: 29916247.

\bibitem{weber18}
B.~Weber, Y.~Nagata, S.~Ketzetzi, F.~Tang, W.~J. Smit, H.~J. Bakker, E.~H.~G.
  Backus, M.~Bonn, and D.~Bonn, ``Molecular insight into the slipperiness of
  ice,'' {\em J. Phys. Chem. Lett.}, vol.~9, no.~11, pp.~2838--2842, 2018.
\newblock PMID: 29741089.

\bibitem{murata15}
K.-i. Murata, H.~Asakawa, K.~Nagashima, Y.~Furukawa, and G.~Sazaki,
  ``\textit{In situ} determination of surface tension-to-shear viscosity ratio
  for quasiliquid layers on ice crystal surfaces,'' {\em Phys. Rev. Lett.},
  vol.~115, p.~256103, Dec 2015.

\bibitem{nagata19}
Y.~Nagata, T.~Hama, E.~H.~G. Backus, M.~Mezger, D.~Bonn, M.~Bonn, and
  G.~Sazaki, ``The surface of ice under equilibrium and nonequilibrium
  conditions,'' {\em Acc. Chem. Res.}, vol.~52, no.~4, pp.~1006--1015, 2019.

\bibitem{tip4p}
W.~L. Jorgensen, J.~Chandrasekhar, J.~D. Madura, R.~W. Impey, and M.~L. Klein,
  ``Comparison of simple potential functions for simulating liquid water,''
  {\em The Journal of Chemical Physics}, vol.~79, no.~2, pp.~926--935, 1983.

\bibitem{Vega2005}
J.~L.~F. Abascal and C.~Vega, ``A general purpose model for the condensed
  phases of water: Tip4p/2005,'' {\em The Journal of Chemical Physics},
  vol.~123, no.~23, p.~234505, 2005.

\bibitem{BerendsenSPCE}
H.~J.~C. Berendsen, J.~R. Grigera, and T.~P. Straatsma, ``The missing term in
  effective pair potentials,'' {\em The Journal of Physical Chemistry},
  vol.~91, no.~24, pp.~6269--6271, 1987.

\bibitem{JorgensenTIP5P}
M.~W. Mahoney and W.~L. Jorgensen, ``A five-site model for liquid water and the
  reproduction of the density anomaly by rigid, nonpolarizable potential
  functions,'' {\em The Journal of Chemical Physics}, vol.~112, no.~20,
  pp.~8910--8922, 2000.

\bibitem{water_models}
S.~P. Kadaoluwa~Pathirannahalage, N.~Meftahi, A.~Elbourne, A.~C.~G. Weiss,
  C.~F. McConville, A.~Padua, D.~A. Winkler, M.~Costa~Gomes, T.~L. Greaves,
  T.~C. Le, Q.~A. Besford, and A.~J. Christofferson, ``Systematic comparison of
  the structural and dynamic properties of commonly used water models for
  molecular dynamics simulations,'' {\em Journal of Chemical Information and
  Modeling}, vol.~61, no.~9, pp.~4521--4536, 2021.
\newblock PMID: 34406000.

\bibitem{blazquez22}
S.~Blazquez and C.~Vega, ``Melting points of water models: Current situation,''
  {\em The Journal of Chemical Physics}, vol.~156, no.~21, p.~216101, 2022.

\bibitem{rates}
K.~G. Libbrecht, ``Physical dynamics of ice crystal growth,'' {\em Annual
  Review of Materials Research}, vol.~47, no.~1, pp.~271--295, 2017.

\bibitem{frict}
R.~Rosenberg, ``Why is ice slippery?,'' {\em Physics Today}, vol.~58, no.~12,
  pp.~50--54, 2005.

\bibitem{coex_rev1}
J.~G. Dash, A.~W. Rempel, and J.~S. Wettlaufer, ``The physics of premelted ice
  and its geophysical consequences,'' {\em Rev. Mod. Phys.}, vol.~78,
  pp.~695--741, Jul 2006.

\bibitem{VegaIce}
J.~L.~F. Abascal, E.~Sanz, R.~García~Fernández, and C.~Vega, ``A potential
  model for the study of ices and amorphous water: Tip4p/ice,'' {\em The
  Journal of Chemical Physics}, vol.~122, no.~23, p.~234511, 2005.

\bibitem{CondeMelting}
M.~M. Conde, M.~Rovere, and P.~Gallo, ``High precision determination of the
  melting points of water tip4p/2005 and water tip4p/ice models by the direct
  coexistence technique,'' {\em The Journal of Chemical Physics}, vol.~147,
  no.~24, p.~244506, 2017.

\bibitem{transport_water1}
M.~A. González and J.~L.~F. Abascal, ``The shear viscosity of rigid water
  models,'' {\em The Journal of Chemical Physics}, vol.~132, no.~9, p.~096101,
  2010.

\bibitem{transport_sanz}
P.~Montero~de Hijes, E.~Sanz, L.~Joly, C.~Valeriani, and F.~Caupin, ``Viscosity
  and self-diffusion of supercooled and stretched water from molecular dynamics
  simulations,'' {\em The Journal of Chemical Physics}, vol.~149, no.~9,
  p.~094503, 2018.

\bibitem{deKonig}
I.~de~Almeida~Ribeiro and M.~de~Koning, ``Non-newtonian flow effects in
  supercooled water,'' {\em Phys. Rev. Research}, vol.~2, p.~022004, Apr 2020.

\bibitem{mcquarrie76}
D.~A. McQuarrie, {\em Statistical Mechanics}.
\newblock New York: Harper \& Row, 1976.

\bibitem{diag_eta}
P.~J. Daivis and D.~J. Evans, ``Comparison of constant pressure and constant
  volume nonequilibrium simulations of sheared model decane,'' {\em The Journal
  of Chemical Physics}, vol.~100, no.~1, pp.~541--547, 1994.

\bibitem{LAMMPS}
A.~P. Thompson, H.~M. Aktulga, R.~Berger, D.~S. Bolintineanu, W.~M. Brown,
  P.~S. Crozier, P.~J. in~'t Veld, A.~Kohlmeyer, S.~G. Moore, T.~D. Nguyen,
  R.~Shan, M.~J. Stevens, J.~Tranchida, C.~Trott, and S.~J. Plimpton,
  ``{LAMMPS} - a flexible simulation tool for particle-based materials modeling
  at the atomic, meso, and continuum scales,'' {\em Comp. Phys. Comm.},
  vol.~271, p.~108171, 2022.

\bibitem{nhchains}
G.~J. Martyna, M.~L. Klein, and M.~Tuckerman, ``Nosé–hoover chains: The
  canonical ensemble via continuous dynamics,'' {\em The Journal of Chemical
  Physics}, vol.~97, no.~4, pp.~2635--2643, 1992.

\bibitem{nhchains2}
G.~J. Martyna, D.~J. Tobias, and M.~L. Klein, ``Constant pressure molecular
  dynamics algorithms,'' {\em The Journal of Chemical Physics}, vol.~101,
  no.~5, pp.~4177--4189, 1994.

\bibitem{p3m}
R.~Hockney and J.~Eastwood, {\em Computer Simulation Using Particles}.
\newblock CRC Press, 1988.

\bibitem{diff_expT298}
K.~Krynicki, C.~D. Green, and D.~W. Sawyer, ``Pressure and temperature
  dependence of self-diffusion in water,'' {\em Faraday Discuss. Chem. Soc.},
  vol.~66, pp.~199--208, 1978.

\bibitem{diff_expT273}
F.~X. Prielmeier, E.~W. Lang, R.~J. Speedy, and H.-D. Lüdemann, ``The pressure
  dependence of self diffusion in supercooled light and heavy water,'' {\em
  Berichte der Bunsengesellschaft für physikalische Chemie}, vol.~92, no.~10,
  pp.~1111--1117, 1988.

\bibitem{viscosity_exp}
K.~R. Harris and L.~A. Woolf, ``Temperature and volume dependence of the
  viscosity of water and heavy water at low temperatures,'' {\em Journal of
  Chemical \& Engineering Data}, vol.~49, no.~4, pp.~1064--1069, 2004.

\bibitem{triple_point}
C.~Vega, J.~L.~F. Abascal, and I.~Nezbeda, ``Vapor-liquid equilibria from the
  triple point up to the critical point for the new generation of tip4p-like
  models: Tip4p/ew, tip4p/2005, and tip4p/ice,'' {\em The Journal of Chemical
  Physics}, vol.~125, no.~3, p.~034503, 2006.

\bibitem{glassformers}
J.~A. Hodgdon and F.~H. Stillinger, ``Stokes-einstein violation in
  glass-forming liquids,'' {\em Phys. Rev. E}, vol.~48, pp.~207--213, Jul 1993.

\bibitem{dielectric_comp}
J.~L. Aragones, L.~G. MacDowell, and C.~Vega, ``Dielectric constant of ices and
  water: A lesson about water interactions,'' {\em The Journal of Physical
  Chemistry A}, vol.~115, no.~23, pp.~5745--5758, 2011.
\newblock PMID: 20866096.

\bibitem{stension_comp}
C.~Vega and E.~de~Miguel, ``Surface tension of the most popular models of water
  by using the test-area simulation method,'' {\em The Journal of Chemical
  Physics}, vol.~126, no.~15, p.~154707, 2007.

\bibitem{comparison_properties}
C.~Vega, J.~L.~F. Abascal, M.~M. Conde, and J.~L. Aragones, ``What ice can
  teach us about water interactions: a critical comparison of the performance
  of different water models,'' {\em Faraday Discuss.}, vol.~141, pp.~251--276,
  2009.

\bibitem{diff_2005}
G.~Guevara-Carrion, J.~Vrabec, and H.~Hasse, ``Prediction of self-diffusion
  coefficient and shear viscosity of water and its binary mixtures with
  methanol and ethanol by molecular simulation,'' {\em The Journal of Chemical
  Physics}, vol.~134, no.~7, p.~074508, 2011.

\end{thebibliography}

\end{document}